\documentstyle[preprint,tighten,aps]{revtex}

\def\Fiissm{\Phi_{i,SSM}}
\def\Fii{\Phi_i}

\def\FiBe{\Phi_{Be}}
\def\FiB{\Phi_{B}}

\def\z{{\text{z}}}
\def\k{{\text{k}}}
\def\s{{\text{s}}}
\def\t{{\text{t}}}

\def\p{{\text{p}}}
\def\lapprox{\mathrel{\mathop
  {\hbox{\lower0.5ex\hbox{$\sim$}\kern-0.8em\lower-0.7ex\hbox{$<$}}}}}
\def\gapprox{\mathrel{\mathop
  {\hbox{\lower0.5ex\hbox{$\sim$}\kern-0.8em\lower-0.7ex\hbox{$>$}}}}}

\begin{document}

%
\title{Helioseismic constraints to the central solar temperature and
 neutrino fluxes}

\author{ B.Ricci $^{1}$,
         V. Berezinsky $^{2}$,
         S.~Degl'Innocenti$^{1,3}$,
         W. A. Dziembowski$^{4}$,
         G.~Fiorentini$^{1,5}$, }

\address{
$^{1}$Istituto Nazionale di Fisica Nucleare, Sezione di Ferrara,
      via Paradiso 12, I-44100 Ferrara, Italy\\
$^{2}$Istituto Nazionale di Fisica Nucleare, Laboratori Nazionali del Gran
     Sasso, I-67010 Assergi (L'Aquila), Italy,\\
$^{3}$Dipartimento di Fisica dell'Universit\`a di Pisa,
       piazza Torricelli 1, I-56100 Pisa, Italy\\
$^{4}$Copernicus Astronomical Center, ul. Bartycka 18, 00716 Warsaw, 
Poland\\
$^{5}$Dipartimento di Fisica dell'Universit\`a di Ferrara,
       via Paradiso 12, I-44100 Ferrara, Italy
        }

\date{May 1997}
\maketitle                 

\begin{abstract}

The central solar temperature $T$ and its uncertainties  are 
calculated in helioseismologically-constrained solar models. From the best 
fit to the convective radius, density at the convective radius and 
seismically determined helium abundance the central temperature 
is found to be $T=1.58 \times 10^7$K, in excellent agreement with 
Standard Solar Models. Conservatively, we estimate that
the accuracy of this determination is
$\Delta T/T=1.4\%$, better than that in SSM. Neutrino fluxes are 
calculated. The lower limit to the boron neutrino flux, obtained with 
maximum reduction factors from all sources of uncertainties,  is 
$2 \sigma$ higher than the flux measured recently by SuperKamiokande. 
\end{abstract}

~\\
~\\

Helioseismology allows us to look into the deep interior of the Sun, probably 
more efficiently than neutrinos (for reviews see 
\cite{Science,Oscill,ARAA,elios,c-d}). The highly  precise
 measurements of 
frequencies and  the tremendous number of measured lines enable us to extract 
the values of sound speed and density inside the Sun with accuracy better 
than $1\%$. Recently it was 
demonstrated that a comparable accuracy 
can be obtained for 
the inner core of the Sun ($R/R_{\odot} < 0.1$) \cite{Dz,eliosnoi}.

Helioseismic data are in agreement with 
recent Standard Solar Model (SSM) calculations,
which use accurate equations of state,
recent opacity tables and include helium and heavier elements diffusion
\cite{BPelios,eliosnoi,BP95,Ciacio,JCD}, see also Ref. \cite{Vau}.
These SSMs yield central 
temperatures 
$T_{SSM}$ which differ from each other by not more than 1\%. However the 
uncertainties in
input parameters, mainly the opacity $\kappa$ and the heavy elements
abundance Z/X,  result in $(\Delta T/T)_{SSM} \approx 1-2\%$.

From helioseismological observations one  cannot determine 
directly temperature of the solar interior, as  one cannot determine
the temperature of a gas from the knowledge of the sound speed unless the 
chemical composition is known. However, it is possible to obtain the range of 
allowed values of the central temperature $T$, by
selecting those solar models which are consistent with 
seismic data.~\footnote{Here and everywhere below $T$ is the central 
temperature.}

In this paper we shall address two related problems: accuracy in the 
determination of the central temperature in helioseismologically-constrained
solar models and neutrino fluxes in these models.

Our calculations are not model-independent, but we shall use in 
principle a wider  class of models in comparison with SSMs, which we call 
helioseismologically-constrained solar models (HCSM). 
These models are based on the same  equilibrium 
and evolution equations as SSMs, but they differ in the choice of 
some input parameters.  We generate this class
of models by using the FRANEC code \cite{Ciacio}
for the SSMs and varying the input 
parameters. 
Each choice of the set of parameters gives some value of $T$.

We obtain the range of allowed values of $T$ by selecting those solar
models which are consistent with seismic data. More specifically, we shall
determine the central temperature $T_{HCSM}$, as that of the model
which gives the best fit to the seismic data and the uncertainties, 
$\Delta T _{HCSM}$,
corresponding to the range spanned by models  consistent with these
data.

As a method of calculation, we shall use the scaling approximation 
of the results of solar models evaluated with FRANEC.
 Namely we assume that 
every physical observable $Q$ (e.g. temperature, photospheric helium abundance, 
radius of convective zone etc) depends on the input parameters $P$ as 
\begin{equation}
\frac {Q} {Q_{SSM}} = \left ( \frac{P}{P_{SSM}}  \right  )^ {\alpha _{Q,\p}}
,
\label{scal}
\end{equation} 
where the subscript SSM is related to the (arbitrarily chosen)
reference SSM.
In this paper we use the new determination
of the coefficients 
$\alpha_{Q,P}$ obtained with  the latest version of FRANEC 
\cite{Ciacio}, which includes diffusion of all elements.

The precise value of the temperature is governed by four quantities $P$:
 the radiative opacity $\kappa$, the fraction  of  heavy elements $Z/X$,
 the astrophysical factor of the p+p reaction $S_{11}$, and
 the solar age $t_\odot$.
If  these four quantities  are rescaled  with respect  to the values used in
the SSM calculation  by a multiplicative factor p$_i =P_i/P_{i,SSM}$
(p$_i$=k, z, s and t respectively), 
the central temperature scales as:
\begin{equation}
\label{Tgen}
T= T_{SSM} \, \k^{0.14}  \, \z^{0.078}  \,  \s^{-0.14} \, \t^{0.084}.
\end{equation}

We shall proceed now  to determine the range of the scaling variables, 
$p_i$,
allowed by seismic observations. 
We remark that the errors on
$S_{11}$ and age $t_{\odot}$ are small (1\% or less),
so that they weakly affect $T$. On the other hand,
the uncertainties on $\kappa$ and Z/X are of the order of  10\%
and their influence on $T$ is important. Furthermore,
these uncertainties do not correspond to clear experimental
or observational errors, rather they are determined 
by judicious comparison among published values. 
As an example, the uncertainty on $\kappa$ can only
be estimated from the comparison
among recent theoretical calculations.

We shall thus concentrate on the two most important scaling variables, k and z. 

We want to emphasize that this not a determination of the 
above-mentioned parameters directly from seismic data, since 
solar models are involved in this evaluation.
Rather we intend to determine those solar models which are consistent
with helioseismology.
Let us also remark that we are restricting  to {\em{uniform}},
although generous, variations of opacity.

As seismic ``observables"  we choose three independent physical 
quantities determined most accuratelly by seismic observations 
(see Ref.\cite{eliosnoi}),
namely the photospheric helium abundance, $Y_{ph}$, the depth of convective 
envelope, $R_b$, and the density, $\rho_b$, at the bottom of convective 
envelope (see the data in Table 1). 
A fourth seismic ``observable", the sound speed at the convective radius
is traditionally considered, e.g. \cite{c-d}. We have not included it in 
our list 
since, as shown in Ref.\cite{elios}, it is not an independent one.

By numerical experiments, using FRANEC, we found the following 
dependence  on z and k (for 
dependence on the other variables see Table \ref{tab1})  
\begin{mathletters}
\begin{eqnarray}
\label{leggi2b}
R_b&=& R_{b, SSM} \, \z^{-0.046} \, \k^{-0.0084}\\
\label{leggi2c}
\rho_b&=& \rho_{b,SSM} \, \z^{0.47} \, \k^{0.095}\\
\label{leggi2a}
Y_{ph}&=& Y_{ph, SSM} \,  \z^{0.31} \,  \k^{0.61}
\end{eqnarray}
\end{mathletters}

 We remark
that $\rho_b$ and $R_b$ have almost equal restriction power. 
In fact the error in $R_b$ is much smaller 
than that in $\rho_b$ (see Table \ref{tab1}),
but the dependence on the scaling parameters is 
much stronger for $\rho_b$, see Eq.(\ref{leggi2c}).

We also remark that $T$ is mainly 
determined by   $Y_{ph}$. One sees
from Eqs. (\ref{Tgen})  and (\ref{leggi2a}) that to
a good approximation the
dependence of $T$ and $Y_{ph}$
 on z and k is just through $\eta= \z \cdot \k ^2$, 
so that one can express $T$ as  a function of $Y_{ph}$:
\begin{equation}
T_{HCSM}= T_{SSM} \,
\left  ( Y_{ph,\odot} / Y_{ph,SSM} \right ) ^{0.2} \quad ,
\end{equation} 

Let us determine now the values of z and k which give the best fit to all 
three quantities $Q_{\odot,i}$, i.e. $Y_{ph}, R_b$ and $R_b$. For this 
we minimize the function 
\begin{equation}
\chi ^2 (\z,\k) =
\sum _i \left( \frac {Q_{\odot,i} -  Q_i (\z,\k)}
{\Delta Q_{\odot,i}} \right ) ^2  
\quad .
\label{xi}
\end{equation}
The corresponding value of the central temperature $T(\z,\k)$ gives our 
best estimate $T_{HCSM}$.

By  starting with BP95 model (the ``best model with metal and helium diffusion"
of Ref.\cite{BP95}) with a central temperature
$T_{SSM}=1.584 \times 10^7~K$, one arrives at $T_{HCSM}=1.587 \times 10^7~K$,
slightly higher but fully 
 consistent with the former value within the uncertainties of SSMs.
In fact, this shows that BP95  is in good agreement with 
helioseismic data. 
We repeated  the same procedure starting from a few different SSM calculations,
on which enough information is available to us:
FR97 is the ``best'' model
with  He and heavier elements diffusion of
Ref. \cite{Ciacio};  
FR96 is another variant of the same model with the Livermore opacities tables 
calculated
for just 12 elements \cite{eliosnoi};
 JCD is the ``model S'' of Ref. \cite{JCD}.

We remark that BP95 and FR96 adopt the same 12 elements composition 
\cite{IR12}, 
whereas FR97 uses the newest Livermore opacities calculated 
for 19 elements \cite{IR96}.
The values found are all within the range 
$T_{HCSM}=(1.573-1.587) \times 10^7$K,
slightly tighter than that of SSM predictions, 
$T_{SSM}=(1.567-1.584) \times 10^7$K, see also
Fig. 1.

By averaging over all available SSM as  starting 
models we obtain the best estimated seismic temperature
\begin{equation}
T_{HCSM}=1.58 \times 10^7~K \, .
\label{Thcsm}
\end{equation}

The range of acceptable temperatures is determined 
by those values of z and k such that {\em each} $Q(\z,\k)$ 
is in agreement with helioseismology 
within the  estimated uncertainty.
By using BP95
as a reference SSM,
the allowed  domain of z and k
is shown in Fig. 2,  where we also show the limiting temperatures. 
The resulting uncertainty
is $\pm 1\%$.

A conservative estimate of $(\Delta T/T)_{HCSM}$ is obtained
by adding the spread due to  starting with different 
SSMs. The result is shown together with $(\Delta T/T)_{SSM}$
estimated with the same procedure, assuming a 10\% uncertainty in 
$\kappa$ and Z/X:
\begin{equation}
(\Delta T/T)_{HCSM}= \pm 1.4\%,\,\, 
(\Delta T/T)_{SSM}= \pm 2.7\%
\label{deltal}
\end{equation}  

We remark that  Eq. (\ref{deltal}) for both uncertainties
is derived using a similar and conservative approach,
of {\em adding up linearly all error sources}.

Should one add errors in quadrature (so that e.g. 
$\Delta Y_{ph}/Y_{ph}=1.4\%$, see \cite{eliosnoi}), it gives
\begin{equation}
(\Delta T/T)_{HCSM}= \pm 0.5\%,\,\,\,(\Delta T/T)_{SSM}= \pm 1.7\%  
\label{deltaq}
\end{equation}

Fig. 1 summarizes
what we have found so far: 
 the central temperatures in
HCSMs agree very well with ones in SSMs, though the uncertainties in
the former models are  smaller.

Finally, we shall discuss a question which a concerned reader certainly 
 must ask: why do we not use the sound speed profile as a constraining 
condition? Indeed, in the region $0.2 < R/R_{\odot} <0.6$ the accuracy on
 the isothermal sound speed squared 
$U={\cal P}/\rho$,
where ${\cal P}$ is pressure, is better than 0.5\%. Why this tremendous 
accuracy, taken as a function of distance, does not give the strongest 
constraints?

Let us look to this problem quantitatively.
At each fractional  distance $x=R/R_{\odot}$ we parametrize $U$ as:
\begin{equation}
U(x) =U _{SSM}(x) \, \z^{\alpha _{\z} (x)}  \, \k ^{\alpha _{\k} (x) } 
\, \s^{\alpha _{\s} (x) }
\end{equation}
The calculated coefficients, $\alpha_{\p}(x) \equiv d \ln U/d \ln \p$,
are plotted in Fig. 3. 
For each parameter $\p$ the 
excluded value of $\Delta \p$ is given by
\begin{equation}
 \Delta \p  \geq \frac{\Delta U}{U \, \alpha_{\p}} \, .
\label{excl}
\end{equation}
In case of $Z/X$, for example, the maximum $\alpha _\z$ is 0.05 (see Fig.3) 
and for $\Delta U/U \approx 0.005$ we can exclude  variations 
in $Z/X$ of order of 10\% or more, which does not improve our knowledge.

As noted in Ref. \cite{Tripathy}, a change in $\kappa (\rho,T)$ by
a multiplicative  constant factor  can  largely be compensated by change
in the composition, the sound speed profile remaining approximately the same in 
the intermediate region.
Actually the opacity coefficients are an order of
magnitude smaller than the others,
so that a 10\% uniform variation of $\kappa$, which affects $T$ to the 1.5\%
level [see Eq. (\ref{Tgen})], cannot be excluded by studying 
$U(x)$.

Let us come over to {\em neutrino fluxes}, $\Phi_i$ ($i$=pp, Be, B).
Their dependence  on the central 
temperature $T$ is parametrized  as:
\begin{equation}
\label{lexf}
 \Fii =\Fiissm \left  ( \frac{T}{T_{SSM}} \right ) ^{\beta _i} \, .
\end{equation}
From numerical experiments with FRANEC, we found:
$\beta _{pp}=-0.92,\,\beta _{Be}=7.9, \, \beta _{B}=18$.

We have determined neutrino fluxes by starting with different SSMs,
renormalizing their predictions to the same temperature $T_{HCSM}=1.58\cdot
10^7$ K
and to the same (updated) nuclear cross sections. The resulting fluxes and
signals, all very close to each other, have been averaged to
determine the HCSM predictions shown in Table \ref{tabsig},
 where the uncertainties
correponding to  $(\Delta T/ T)_{HCSM}=\pm 1.4\%$ are also indicated.        

In order to discuss nuclear physics uncertainties, 
we shortly review the present status  for three 
most important cross-sections. The numbers quoted below refer to $3\sigma$ 
errors.\\
\noindent
a) $^3$He+$^3$He $\rightarrow$ $^4$He + 2p.
This cross-section was measured recently in LUNA experiment \cite{Arpe} at 
energy corresponding to the solar Gamow peak ($\sim 20$ KeV). The 
uncertainty in astrophysical factor is  $\Delta S_{33}/S_{33}  = \pm 6 \%$.\\
\noindent
b) $^3$He+$^4$He $\rightarrow$  $^7$Be + $\gamma$.
 The uncertainty is  $\Delta S_{34}/S_{34}  = \pm 12 \%$.\\
\noindent
c) p+$^7$Be $\rightarrow$ $^8$B  + $\gamma$.
 $S_{17} $
is determined from direct measurements with an error of about
 $ \pm30 \%$  \cite{koonin2,KAV69,FIL83}. The indirect 
measurement \cite{MOTO} gives  $S_{17}$ consistent with
the lower limit of direct measurements, 
although the accuracy of this method 
is unclear \cite{LS,GAIBERT,LSreply}.

The errors in fluxes due to cross-sections are calculated using the 
relations between fluxes and cross-sections \cite{AA,Report},
$\FiBe \propto S_{34} \, S_{33} ^{-1/2} $ and
$\FiB \propto S_{34} \, S_{33}  ^{-1/2} \, S_{17}$.
We remark that the  (3$\sigma$) uncertainty on $\FiBe$  due to that on
 $S_{34}$ is slightly larger than that corresponding to temperature.
The 30\%   uncertainty on
$\FiB$ due to that of $S_{17}$ exceeds that due to the solar temperature.

One concludes, see Table \ref{tabsig}, that ($3\sigma$) nuclear
physics uncertainties are at least as important as the 
(generously estimated 1.4\%) temperature uncertainty.

In the end we shall discuss the problem of boron neutrinos.

The HCSM boron neutrino flux (see Table II) is $18\sigma$ higher than 
than the combined Kamiokande \cite{Kamioka} and SuperKamiokande \cite{SK} flux,
$\Phi_{K}=(2.58 \pm 0.19) \cdot 10^6~cm^{-2}s^{-1}$). However, the 
uncertainties in the predicted flux are of great importance for this 
comparison.

During the last several  years an idea of reconciling  the predicted 
boron neutrino flux with that measured by Kamiokande was widely discussed
\cite{ShSh,Dar,Berez,BFL}. This discussion was inspired 
by the low value of $S_{17}$ measured in indirect experiments \cite{MOTO}
and by a possible decrease of the central temperature $T$ due to collective 
plasma effects (through opacity \cite{Tsyt}) and due to abundance of heavy
elements Z. These effects, when correlated, can result in the agreement 
between predicted and measured B-neutrino flux.

Since in our calculations the opacity and heavy elements abundance are 
constrained by seismic observations, the status of this problem has changed.  
Indeed, diminishing $T_{HCSM}$ by 1.4\% and $S_{17}$ by 30\% one obtains 
from the boron-neutrino flux in HCSM the 
minimum flux $3.24\cdot 10^6~cm^{-2}s^{-1}$, which is $3.5 \sigma$ higher 
than the combined Kamiokande and SuperKamiokande flux. 

If one accepts a more reasonable $1\%$ reduction of $T_{HCSM}$, but  takes 
other 
errors in a correlated way, all diminishing the B-neutrino flux (namely
$S_{17}$ by $30\%$ smaller, $S_{34}$ by $12\%$ smaller and $S_{34}$ by 
$6\%$ higher) we obtain $\Phi_B=2.97\cdot 10^6~cm^{-2}s^{-1}$, i.e. 
$2.1 \sigma$  higher than 
SuperKamiokande flux. Only in case of largest possible correlated errors 
the HCSM flux can be reconciled with the SuperKamiokande data.

Therefore, we  have now three solar-neutrino problems:
(i) the controversy of Homestake and SuperKamiokande results, 
(ii) the Beryllium neutrino problem, as controversy of gallium and 
SuperKamiokande (or Homestake) experiments, and 
(iii) Boron neutrino problem, as it is described above.

In conclusion, the helioseismologically constrained solar models (HCSM) give 
the central solar temperature in excellent agreement with SSMs, and with 
smaller uncertainties. The boron neutrino problem exists in this class of
models.

\acknowledgments
This work started  during an enjoyable meeting at Osservatorio Astronomico 
di Collurania (O.A.C.T.) near Teramo
 and we express our gratitude to the  director, V. Castellani 
for the pleasant atmosphere and for the fruitful
discussions. We also thank the O.A.C.T. administration
for organization efforts. This work was supported in part by the Human 
Capital and Mobility Program of the European Economic Community under 
Contract No. CHRX-CT93-0120 (CG 12 COMA). W.A.D.  gratefully acknowledges the 
financial support from Laboratori Nazionali del Gran Sasso of INFN.

\begin{table}
\caption[aa]{
Seismically determined quantities, Q, and the exponents $\alpha_{Q,P}$ of the 
scaling approximation given by Eq.(1).
}
%
%
%
\begin{tabular}{l c c c c c}
Q & Q$_{\odot}$&$\alpha_{Q,k}$&$\alpha_{Q,z}$&$\alpha_{Q,s}$&$\alpha_{Q,t}$ \\
\hline
$Y_{ph}$ &   0.249 ($1\pm 4.2\%$) & 0.61 &0.31 & 0.14  &0.20 \\
$R_b/R_\odot$  & 0.711 ($1\pm 0.4\%$) &-0.0084 & -0.046 & -0.058 &-0.080 \\
$\rho_b$ [gr/cm$^3$] &  0.192  ($1\pm 3.7\%$) & 0.095 & 0.472 &0.86 & 0.85 \\
\end{tabular}
\label{tab1}
\end{table}

\begin{table}
\caption[abc]{
Predictions for neutrino fluxes and signals in the Cl and Ga detectors 
from SSM and HCSM. Uncertainties corresponding
 to $(\Delta T/T )_{HCSM}=\pm 1.4\%$ 
are shown (first error) together with those from
 nuclear cross sections (second error).
}
\begin{tabular} {ll | c c c c c| c c c c c }
 & &\multicolumn{4}{c}{SSM} && \multicolumn{4}{c}{HCSM}\\
 & & BP95 & FR97 & FR96  & JCD  &&  \\
\hline
$\FiBe$ &$[10^9$/cm$^{2}$/s$]$ & 5.15 & 4.49 & 4.58 & 4.94 &&  4.81$\pm 0.53 \pm 0.59$ & \\
 $\FiB$ &$[10^6$/cm$^{2}$/s$]$ & 6.62 &  5.16& 5.28 & 5.87 && 5.96$\pm 1.49 \pm 1.93$ &  \\
Cl &$[$SNU$]$& 9.3 & 7.3 & 7.5 & 8.2 && 8.4$\pm 1.9 \pm 2.2$   &\\
Ga &$[$SNU$]$&137 &128& 129& 132 &&  133$\pm 11 \pm 8$ &\\
\end{tabular}
\label{tabsig}
\end{table}

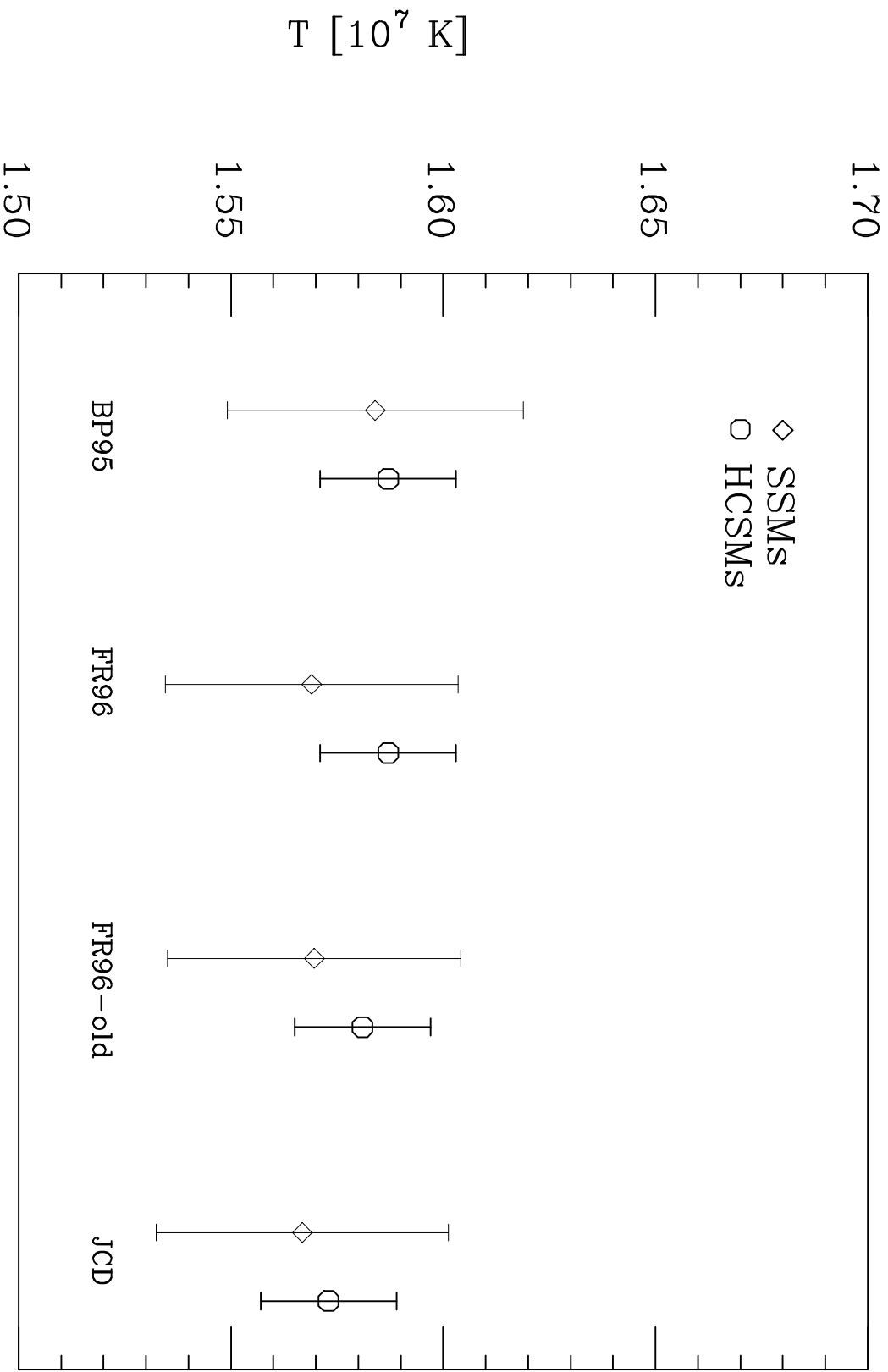
\begin{figure}
\caption[b]{
For a few recent SSM calculations we present :
a) the predicted central solar temperatures  $T_{SSM}$  (diamonds)
with the conservative uncertainty  (thin bars);
b) the  values $T_{HCSM}$ derived  by  best fit of  helioseismic 
observables at the convective radius (circles)  
with the uncertainties (bars) calculated in the same way as for SSM's.
Labels indicating solar models are defined in the text.
}
%
\label{fig1}
\end{figure}

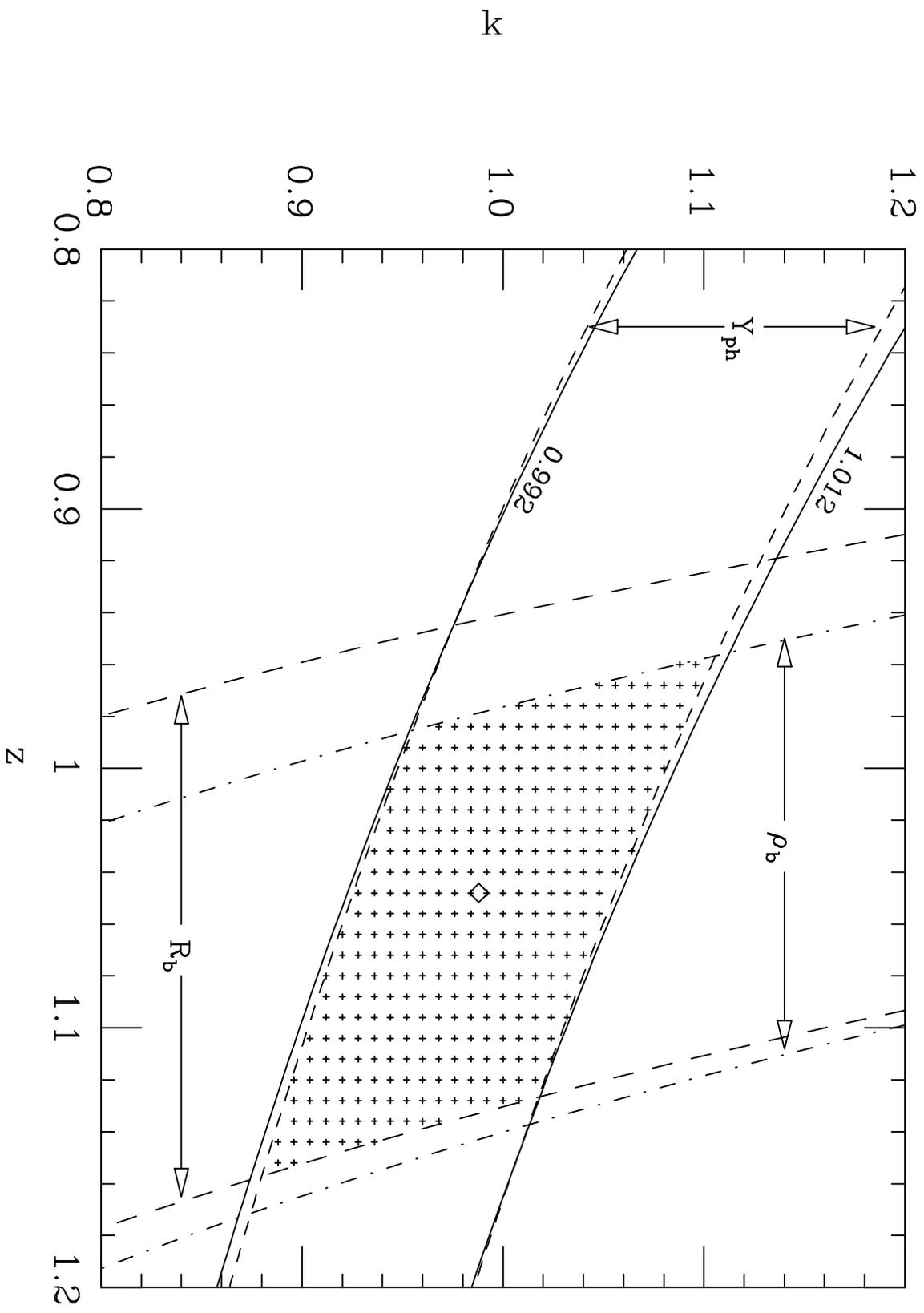
\begin{figure}
\caption[bb]{
In the (z,k) plane we show the position of the best-fit HCSM
(diamond) and of the models consistent with helioseismic data (crosses) 
obtained by  using BP95 as a starting model. The curves labelled by 
$\rho_b$, $Y_{ph}$ and $R_b$ show the helioseismic constraints due to 
these quantities. Curves labelled by the numbers are isotemperature 
curves in $kz$ plane. Numbers are values of $T/T_{SSM}$.
}
\label{fig2}
\end{figure}

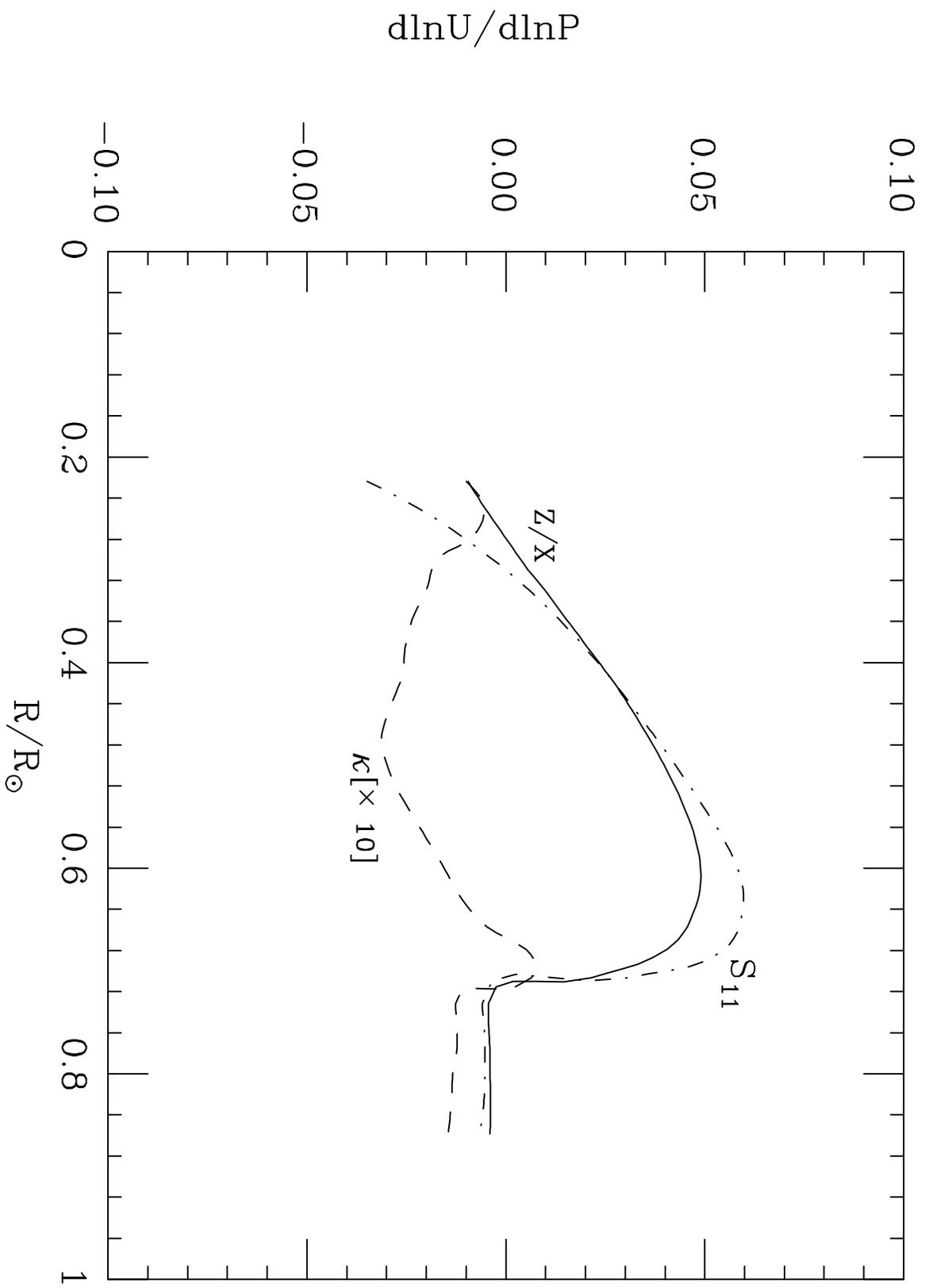
\begin{figure}
\caption[bbbb]{
Logaritmic derivates of the isothermal sound speed squared $U$ with respect
to $S_{11}$ (dot-dashed line), opacity (dashed line) and Z/X (full line). Note
that the opacity coefficient have been multiplied by  a factor 10.
}
\label{figalpha}
\end{figure}

\end{document}